\documentclass[twoside,journey]{IEEEtran}
\usepackage{makecell}
\usepackage{hyperref}
\usepackage{array}
\usepackage{graphicx,amssymb,amsmath}
\usepackage{multicol}
\usepackage[noadjust]{cite}
\usepackage{setspace}
\usepackage{subfigure}
\usepackage{graphicx}
\usepackage{float}
\usepackage {url}
\usepackage{stfloats}
\usepackage{amsthm,pifont}
\usepackage{cases,subeqnarray}
\usepackage{bm,multirow,bigstrut}
\usepackage{amsmath, amsthm, amssymb}
\usepackage{textcomp}
\usepackage{latexsym,bm}
\usepackage{booktabs}
\usepackage{xcolor}
\usepackage{mathtools}
\usepackage{dsfont}
\usepackage{extarrows}
\usepackage{epsfig}
\usepackage{epstopdf}
\usepackage{pifont}
\usepackage{flushend}
\usepackage[noend]{algpseudocode}
\usepackage{algorithmicx,algorithm}
\usepackage{booktabs}
\usepackage{tabularx}
\usepackage{graphicx,tabularx,booktabs,hyperref}
\usepackage{booktabs}      
\usepackage{tabularx}      
\usepackage[table]{xcolor} 
\usepackage{graphicx}      
\usepackage{hyperref} 
\definecolor{linkInk}{HTML}{2F4858} 
\hypersetup{colorlinks=true, urlcolor=linkInk}

\definecolor{rowBlueL}{HTML}{F1F6FF}
\definecolor{rowBlueD}{HTML}{E1ECFF}
\definecolor{rowGreenL}{HTML}{F2FFF4}
\definecolor{rowGreenD}{HTML}{E4F9E7}
\definecolor{rowOrangeL}{HTML}{FFF8ED}
\definecolor{rowOrangeD}{HTML}{FFF0DD}
\definecolor{tableHeaderGray}{gray}{0.88} 
\definecolor{tableRowGray}{gray}{0.96}     
\definecolor{linkBlue}{RGB}{65, 105, 225} 
\theoremstyle{plain}

\theoremstyle{plain}

\usepackage{amsmath}

\IEEEoverridecommandlockouts
\begin{document}

\title{Unleashing Tool Engineering and Intelligence for Agentic AI in Next-Generation Communication Networks}
\author{Yinqiu Liu, Ruichen Zhang, Dusit Niyato,~\textit{Fellow, IEEE}, Abbas Jamalipour,~\textit{Fellow, IEEE}, \\Trung Q. Duong,~\textit{Fellow, IEEE}, and Dong In Kim,~\textit{Life Fellow, IEEE}

\thanks{Y. Liu, R. Zhang, and D. Niyato are with the College of Computing and Data Science, Nanyang Technological University, Singapore (E-mail: yinqiu001@e.ntu.edu.sg, ruichen.zhang@ntu.edu.sg, and dniyato@ntu.edu.sg).}
\thanks{A. Jamalipour is with the School of Electrical and Computer Engineering, University of Sydney, Australia, and with the Graduate School of Information Sciences, Tohoku University, Japan (E-mail: a.jamalipour@ieee.org).}
\thanks{T. Q. Duong is with the Faculty of Engineering and Applied Science, Memorial University, Canada, and with the School of Electronics, Electrical Engineering and Computer Science, Queen’s University Belfast, U.K., and also with the Department of Electronic Engineering, Kyung Hee University, South Korea (E-mail: tduong@mun.ca).}
\thanks{D. I. Kim is with the Department of Electrical and Computer Engineering, Sungkyunkwan University, South Korea (E-mail: dongin@skku.edu).}
\vspace{-1cm}
}
\maketitle

\begin{abstract}
Nowadays, agentic AI is emerging as a transformative paradigm for next-generation communication networks, promising to evolve large language models (LLMs) from passive chatbots into autonomous operators.
However, unleashing this potential requires bridging the critical gap between abstract reasoning and physical actuation, a capability we term tool intelligence. In this article, we explore the landscape of tool engineering to empower agentic AI in communications. 
We first analyze the functionalities of tool intelligence and its effects on communications. 
We then propose a systematic review for tool engineering, covering the entire lifecycle from tool creation and discovery to selection, learning, and benchmarking.
Furthermore, we present a case study on tool-assisted uncrewed aerial vehicles (UAV) trajectory planning to demonstrate the realization of tool intelligence in communications. 
By introducing a teacher-guided reinforcement learning approach with a feasibility shield, we enable agents to intelligently operate tools.
They utilize external tools to eliminate navigational uncertainty while mastering cost-aware scheduling under strict energy constraints. 
This article aims to provide a roadmap for building the tool-augmented intelligent agents of the 6G era.
\end{abstract}


\IEEEpeerreviewmaketitle
\vspace{-0.2cm}
\section{Introduction}
\label{sec:introduction}
Recent breakthroughs in large language models (LLMs) are demonstrating a profound impact on communication intelligence, with proven successes in complex tasks (e.g., resource allocation \cite{10582827} and channel prediction \cite{10582829}), where traditional automation is increasingly strained \cite{10582827}. Building on this momentum, the communications sector is rapidly advancing from standalone LLMs toward sophisticated agentic AI systems. By embedding LLMs as the core reasoning engine within a framework of perception, memory, and action components, agentic AI emerges as an intelligent entity \cite{11162291}. This agent-based paradigm is pivotal, as it enables the autonomous pursuit of complex and multi-step communication tasks with minimal human supervision, directly aligning with the vision for fully intelligent sixth-generation and beyond communications \cite{11162291}.

The key to unlocking this agentic potential is \textit{tool intelligence}: the capability for an agent to effectively interact with real-world infrastructures through the use of external tools. Tool intelligence bridges the gap from abstract reasoning to concrete, real-world actuation, moving beyond isolated task automation to reshape workflows. For instance, upon detecting user complaints about poor connectivity, HPE Marvis\footnote{Available at: https://www.hpe.com/sg/en/marvis-ai.html} can proactively initiate a diagnostic workflow. Instead of merely generating alerts, it invokes a network telemetry analytics tool to correlate user-experience data with real-time performance metrics, a configuration inspection utility to identify anomalies across underlying switches, and a traffic analysis module to trace packet loss patterns. Once a misconfigured aggregation switch is identified as the root cause, the system can autonomously generate and deploy the corrective configuration, completing the loop from perception to action.

As the application of agentic AI expands, tool intelligence is developing rapidly. In the industry, mainstream platforms, such as OpenAI's GPT and Google's Gemini, have seamlessly integrated foundational tools such as code interpreters and web browsers. They have also fostered a burgeoning ecosystem for third-party tools, from general utilities like \textit{WolframAlpha} for complex computation to platform-specific plugins for services like \textit{Expedia} and \textit{Zapier}. In academia, foundational research has demonstrated how LLMs can be trained to master thousands of real-world application programming interface (API) \cite{qin2023toolllm}, learn to self-correct tool usage \cite{cai2023tool}, and even teach themselves to use tools they have never seen before \cite{11162291}. 
By connecting an agent's reasoning to external capabilities, tools profoundly enhance its power, providing many benefits. 
\begin{itemize} 
\item \textbf{Grounding and Actuation:} Tools provide the essential interface for environmental grounding and real-world actuation, enabling agents to query real-time states and execute control commands. For instance, an agent can use a \textit{Prometheus} client\footnote{Available at: https://github.com/pab1it0/prometheus-mcp-server} to query live network telemetry or invoke a \textit{gNMI} API\footnote{Available at: https://github.com/openconfig/gnmi} to reconfigure a router.
\item \textbf{Specialized Expertise:} Tools grant access to a vast library of domain-specific algorithms that an LLM cannot accurately reproduce from its pre-trained knowledge. This includes invoking a \textit{MATLAB} simulator\footnote{Available at: https://github.com/neuromechanist/matlab-mcp-tools} for complex beamforming calculations or calling a retriever to fetch 3GPP standards for protocol compliance.
\item \textbf{Scalability \& Modularity:} Agents can decompose complex problems into a manageable chain of sub-tasks via tools, fostering a modular and scalable approach to automation. For example, they can orchestrate a workflow involving a fault detection tool, a root-cause analysis tool, and a trouble-ticketing tool to manage the entire lifecycle of a network incident. 
\end{itemize}

{\color{black}
Nonetheless, unleashing the full potential of tool intelligence in agentic AI-empowered communications requires filling the following gaps. 
First, our investigation reveals that mainstream LLM tool ecosystems are dominated by user- and web-centric applications\footnote{Representative tools can be found at: https://mcp.so/}. 
In contrast, there is a significant scarcity of professional tools specifically designed for communications and networking.
Second, the unique characteristics of communication networks impose much stricter requirements on tool operations. 
For instance, the additional latency and computational overhead incurred by tool invocation, while acceptable for a web search, can be prohibitive for real-time communication functions, such as vehicle-to-vehicle transmissions. 
Finally, the deterministic and high-stakes nature of communication networks means that a tool failure, or an LLM's probabilistic error in managing tools, could lead to service degradation/outages \cite{10890828}.

In this article, we explore tool engineering and intelligence for agentic AI in next-generation communication networks. Specifically, we first analyze the importance of tool intelligence by examining the core limitations of LLMs, followed by the role tools play within the agentic AI framework and how they transform communication systems. Moreover, we review the crucial techniques to implement this vision in communication scenarios, a process that we term \textit{tool engineering}. Finally, we conduct a case study on optimizing tool-assisted uncrewed aerial vehicle (UAV) communications. Our goal is to equip readers with a systematic understanding of how to integrate existing tools, optimize tool usage, and create new tools to build next-generation wireless systems.

The main contributions of this paper are as follows.
\begin{itemize}
    \item \textit{To the best of our knowledge, this is the first work to comprehensively explore tool intelligence in next-generation communications.} We first identify the motivation for tool intelligence, then describe the foundational role of tools within agentic AI frameworks. In addition, we analyze the transformative effects of tool intelligence on communication functionality and efficiency.
    \item We provide a systematic review of tool engineering, which we define as the set of key techniques required to realize tool intelligence in communication networks. Particularly, our review highlights five core engineering aspects, namely tool creation, discovery, selection, learning, and standard \& benchmarking.
    \item We present a case study on realizing tool intelligence in UAV communications. We propose a teacher-guided reinforcement learning framework to enable tool-assisted trajectory planning, optimizing the agent's capability to intelligently schedule tool activations while strictly adhering to resource constraints.
\end{itemize}
\begin{figure*}[tbp!]
  \centering
  \includegraphics[width=0.8\textwidth]{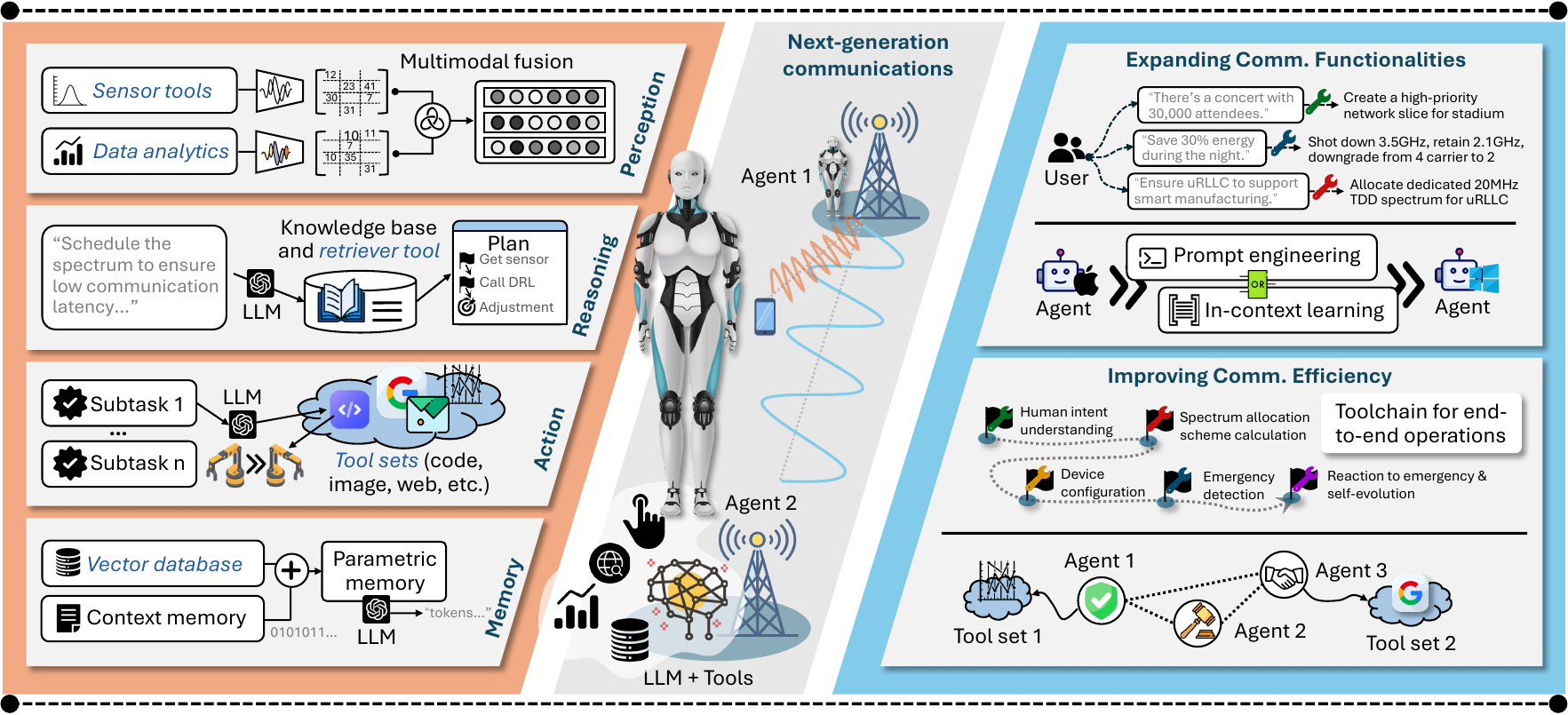} 
  \vspace{-0.2cm}
\caption{(left): The general agentic AI framework, illustrating how tool intelligence is deeply integrated into all four core components: perception (e.g., sensor and data analytics tools), reasoning (e.g., knowledge base and retriever tools), action (tool sets), and memory (e.g., vector databases). (right): The effects of tool intelligence.}
  \vspace{-0.3cm}
  \label{Figure-1}
\end{figure*}

\section{Agentic AI with Tool Intelligence for Next-Generation Communications} \label{sec:preliminaries} 

\subsection{Motivation for Tool Intelligence}
Although conventional AI agents have existed for decades (e.g., deep reinforcement learning (DRL) agents \cite{10582827}), the recent integration of LLMs as core reasoning engines has dramatically expanded their cognitive capabilities \cite{jiang2025tutorial}. 
Nonetheless, LLM-based agents are constrained by three limitations inherent in their LLM cores, which directly motivate the need for external tools.

The first challenge is regarding knowledge grounding and reliability. Since agents' LLMs are trained on static data, they often operate with outdated information and without access to real-time, communication-specific knowledge. This leads to the risk of error and hallucination, where the agent might generate factually incorrect or nonsensical content. Therefore, a communication agent must base its decisions on up-to-date and domain-specific data, creating a critical need for tools that connect to external knowledge sources, such as real-time network monitors, industrial standards (e.g., 3GPP), and scientific literature \cite{jiang2025tutorial}.

Second, agents face the challenge of translating reasoning into action. LLMs are reasoning and content-generation engines, which cannot directly execute a communication command, run a complex simulation, or perform a mathematical optimization. Therefore, tools are indispensable actuators that bridge the gap between the agent's cognitive processes and the physical or virtual network infrastructure. This is especially critical in the communications field, where operational environments are heterogeneous ecosystems of multi-vendor equipment, diverse protocols, and proprietary APIs that require precise, structured commands for interaction \cite{tong2025wirelessagent}.

Third, a fundamental trade-off exists between an LLM's generalist knowledge and the specialized precision required for many communication tasks. The primary strength of an LLM is its vast, pre-trained knowledge, providing generalizability and zero-shot capabilities \cite{jiang2025tutorial}. However, LLMs still exhibit inherent limitations in deep logical inference, precise mathematical computation, and the kind of high-fidelity simulation essential for network engineering. Consequently, when faced with a complex task like optimizing transmission power, an LLM will rarely match the performance of highly optimized, domain-specific algorithms. 

\vspace{-0.3cm}
\subsection{Tools in Agentic AI} 
A modern agentic AI framework typically consists of four core components, namely perception, reasoning \& planning, action, and memory \& evolution, thereby implementing a continuous cycle of autonomous operation \cite{jiang2025tutorial, tong2025wirelessagent}. 
Particularly, as illustrated in Fig. \ref{Figure-1}(left), tool intelligence plays a critical role in all components.
\begin{itemize} 
    \item \textbf{Perception:} Perception refers to the agent's ability to receive and interpret environmental states. In communication networks, this component involves processing diverse data modalities. Hence, sensors and data analytics software tools are called to acquire and parse raw environmental data, creating the situational awareness upon which all decisions are based.
    \item \textbf{Reasoning \& Planning:} This is the cognitive core of an agent, often driven by an LLM, which analyzes perceived information and formulates a multi-step plan. Tool intelligence is critical in this component. The agent may invoke a knowledge base retriever to fetch 3GPP standards or leverage knowledge graphs to improve reasoning capabilities. These tools provide the specialized and timely information the LLM lacks, leading to a more robust and factually grounded plan. 
    \item \textbf{Action:} Based on the generated plan, the agent executes actions by interacting with its environment. This component is the most explicit application of tools, which act as the agent's ``actuators." For a communication agent, this means calling a tool, such as a resource allocation algorithm, a Python script to reconfigure network slices, or a \textit{gNMI} API\footnote{Available at: https://github.com/openconfig/gnmi} to reconfigure a router. 
    \item \textbf{Memory \& Evolution:} This refers to the agent's capacity to learn and adapt over time. Tools are the mechanisms that enable this process, allowing the agent to read from and write to external memory modules, such as vector databases or files. This self-reflection mechanism allows it to refine its capabilities, leading to continuous improvement in dynamic communication networks. 
\end{itemize} 

\subsection{Effects of Tool-enhanced Agentic AI in Communications}

\subsubsection{Expanding Communication Functionalities}
An agent's LLM core can be fused with a growing suite of specialized tools \cite{qin2023toolllm}. 
This is particularly transformative for the communications domain, which relies on a vast library of professional knowledge and complex mathematical calculations (e.g., channel models in 3GPP or beamforming simulations). 
External tools directly compensate for LLMs' inherently limited domain-specific knowledge and mathematical reasoning capabilities \cite{jiang2025tutorial}. 
Furthermore, a single agent can achieve broad generalization: its capabilities are no longer defined by its LLM's internal training, but by the diverse set of tools it can orchestrate (see Fig. \ref{Figure-1}(right)).
}

\subsubsection{Improving Communication Efficiency}
Compared with traditional single-purpose methods, agents achieve superior efficiency in communication management through intelligent tool chain orchestration. 
As shown in Fig. \ref{Figure-1}(right), an agent can autonomously execute the entire spectrum management lifecycle. 
This involves orchestrating a tool chain that invokes a traffic prediction tool to forecast network-layer demand. 
In parallel, it activates a spectrum sensing tool to identify the current physical-layer supply. 
The agent's LLM core then fuses these statistics and activates an optimization tool to calculate the optimal allocation policy. 
By unifying these traditionally separate tools, the agent minimizes overhead and the potential for cumulative errors.

Furthermore, tools facilitate the realization of collective intelligence, where multiple agents, each possessing distinct and specialized tool sets, collaborate to manage communications. 
As illustrated in Fig. \ref{Figure-1}(right), a multi-agent spectrum management system comprises three collaborative agents with different roles and tool sets. 
Upon detecting an emergency, a public safety agent initiates a negotiation.
A regulatory agent then invokes its policy-checking tool to validate the request against established policies. 
This, in turn, enables a commercial agent to use its optimization tool to calculate an optimal block of spectrum while minimizing its own service disruption. 
This tool-driven collaboration among specialized agents results in a context-aware reallocation that is far more resilient than monolithic, pre-programmed systems.

\begin{figure*}[tbp!]
  \centering
  \includegraphics[width=0.8\textwidth]{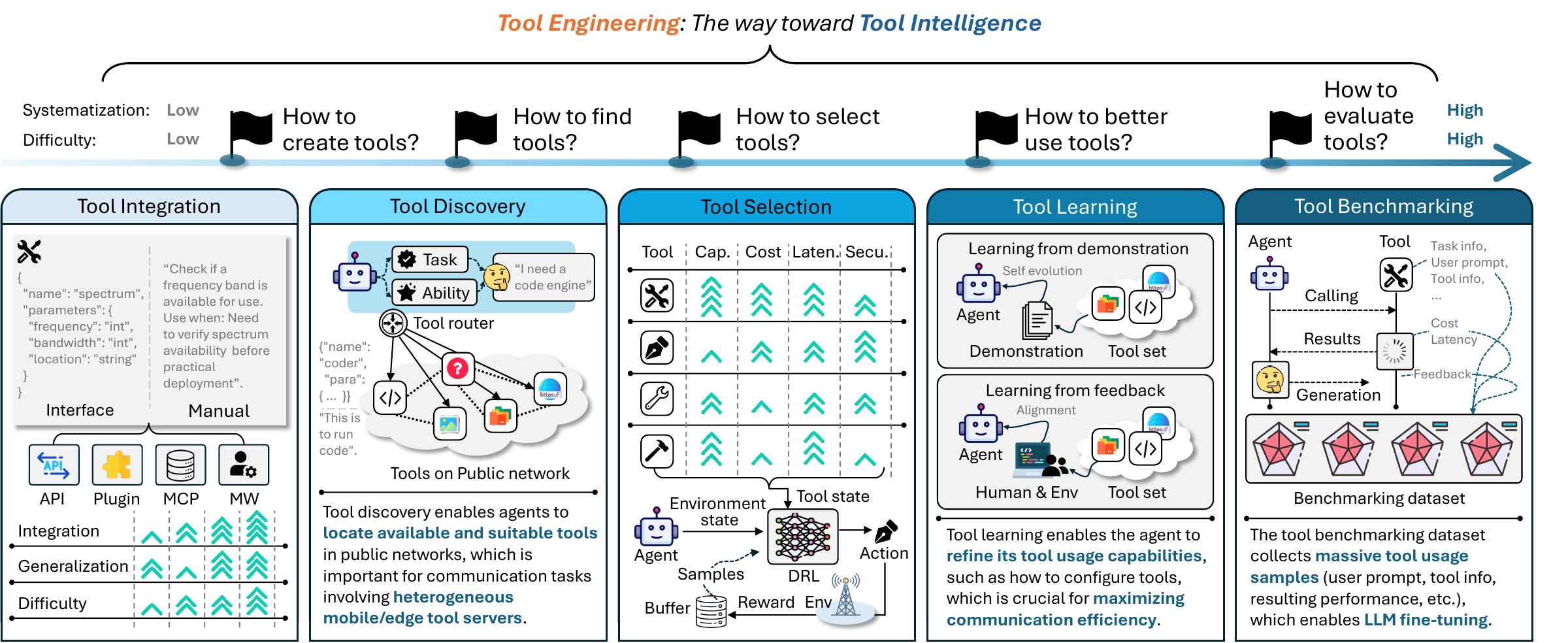} 
  \vspace{-0.25cm}
  \caption{The major aspects of tool engineering. The complexity and systematization increase from left to right, addressing more and more advanced questions. In this way, a complete ecosystem of tool-augmented agentic AI for communications can be implemented.}
  \vspace{-0.3cm}
  \label{Fig2}
\end{figure*}

\section{Tool Engineering: Realizing Tool Intelligence in Communications} \label{sec:enhancing_tools} 



\subsection{Tool Creation and Integration} 
To create a tool for LLM-based agents, two components are essential, i.e., a well-defined interface and a comprehensive natural language description. 
As illustrated in Fig. \ref{Fig2}, the interface provides a structured way for the agent to call the tool with specific parameters and receive the output. 
Equally important is the natural language description, which serves as the ``manual'' for the agent's LLM. 
This description must clearly articulate the tool's purpose, its required input parameters (including their data types and formats), and the structure of its output. 
For example, a beamforming tool might be described as ``Function \texttt{calculate\_beamforming\_weights(location, channel\_state)}''. 
By parsing these descriptions, the agent can understand what tools are available in its arsenal and how to call them correctly \cite{qin2023toolllm}. 

Multiple integration paradigms have emerged. As illustrated in Fig. \ref{Fig2}, direct API integration is the most fundamental, where developers expose REST, gRPC, or Python-callable functions and provide schemas manually to the LLM. Similarly, plugins wrap APIs inside a platform-specific manifest (e.g., ChatGPT or Gemini). 
Although widely adopted, these approaches can be rigid and only suitable for offline usage. 
In contrast, the model context protocol (MCP) \cite{mcpzero} generalizes tool integration by defining a model-agnostic protocol. 
The tools are registered on an MCP server and automatically declare their capabilities in a standardized schema. 
An MCP-compatible client can dynamically discover and invoke them, thereby reducing the integration costs while embedding minimal-privilege and audit controls. 
Finally, middleware, such as LangChain and LangGraph\footnote{Available at: https://docs.langchain.com/}, acts as an orchestration layer that operates APIs, plugins, or MCP endpoints, providing higher-level functionality such as multi-tool reasoning and error handling.

\subsection{Tool Discovery in Open Networks} 
6G and future communications are envisioned as open and distributed ecosystems where new services, functions, and corresponding tools may become available on the fly. 
An agent operating in such an environment cannot rely solely on a fixed, pre-configured toolkit. 
This necessitates a mechanism for automatic tool discovery, allowing agents to dynamically identify and incorporate new tools as they emerge in the network. 
Recent proposals like MCP-zero \cite{mcpzero} aim to address this challenge by creating a decentralized framework for tool registration and discovery. 
Specifically, network functions or services can broadcast their availability and expose their callable tools via a standardized protocol. 
An AI agent can then query this distributed registry to find tools that match its current task requirements. 
For instance, an agent tasked with optimizing a new low-latency service could discover a nearby specialized DRL optimization tool. 

\vspace{-0.2cm}
\subsection{Tool Selection} 

\subsubsection{Static Binding \& Simple Tool Selection} 
The most basic form of tool selection involves static binding or simple keyword matching. In this approach, each tool is hard-coded to a particular type of task, or the agent selects a tool based on a rudimentary analysis of the query's keywords. For example, any user prompt containing the words ``channel quality'' might automatically trigger a tool that queries the network's CSI database. Although simple and predictable, this method is highly inflexible. It fails to account for the nuances of user intent or task context, and it cannot handle situations where multiple tools could potentially address a problem. 

\subsubsection{Dynamic \& Context-aware Tool Selection} 
To perform dynamic tool selections, the agent analyzes the full context of the task, including the specific user intent, interaction history, and the current network state. 
Concurrently, it checks its available toolset, parsing the functionality and description of each tool.
Crucially, for a communication agent, this analysis also incorporates the practical operational metrics of the tools themselves, such as their latency, robustness, and resource consumption, factors that present distinct trade-offs.
This allows the agent to formulate a comprehensive set of KPIs. 
The agent can then leverage LLM's analytical ability to reason about these trade-offs and select the most suitable tool \cite{10.1145/3649476.3658784}. 
Alternatively, as the modeling is highly complex, DRL is also typically used to train a sophisticated selection policy.

\subsubsection{Tool Chain \& Orchestration} 
Many complex communication tasks cannot be solved with a single tool and thus require a sequence of actions, called tool chains.
The agent can act as an intelligent planner, decomposing a high-level goal into a multi-step workflow and orchestrating a chain of tool invocations \cite{agentorc}.
This can range from a simple diagnostic workflow (e.g., monitoring $\rightarrow$ inspection tool $\rightarrow$ configuration) to complex, hierarchical multi-agent systems where a central planning agent delegates tasks to specialized sub-agents with their own tools \cite{agentorc}.

\vspace{-0.3cm}
\subsection{Tool Learning} 
The intelligence of tool usage is not just calling predefined tools but learning to use them better over time, a concept known as tool learning.
This involves the agent refining its ability to select and operate tools based on experience and feedback. 
Through a trial-and-error process, often guided by DRL \cite{10582827}, an agent can learn the subtle nuances of a tool's behavior. 
For instance, it might learn that a particular optimization algorithm performs best under certain channel conditions or that a diagnostic tool is unreliable during peak network load. 
Moreover, this learning process can also be self-driven.
For example, SPORT \cite{SPORT} proposes an iterative framework, where agents optimize tool usage autonomously through step-wise preference tuning. 
After executing a tool chain, the agent can observe the network's response to assess the outcome. 
If the result was suboptimal, the agent can store this experience in its memory and adjust its planning process to avoid making the same mistake in the future \cite{SPORT}. 
This capability is vital in dynamic communication networks, allowing an agent to continuously refine its strategies. 

\vspace{-0.2cm}
\subsection{Tool-Usage Standard, Protocols, and Benchmarking}
Standardization is paramount for scaling the tool ecosystems of communication agents. Although no single universal standard exists yet, the ecosystem is coalescing around several key approaches, such as OpenAI's API and MCP. For tool assessment, early benchmarks, e.g., API-Bank \cite{APIbank} and ToolBench\footnote{Available at: https://github.com/OpenBMB/ToolBench} focus on evaluating an agent's ability to handle a massive number of diverse APIs. More recent benchmarks provide deeper, more nuanced evaluations by focusing on specific real-world challenges. For example, ToolQA tests an agent's ability to answer questions that can only be solved by using tools to query external knowledge, thereby minimizing the influence of the LLM's internal memory. Concurrently, Tool Playground \cite{10890828} evaluates an agent's ability to handle tool failures or correct invalid parameters. Moreover, General Tool Agents (GTA) \cite{GTA} construct a real-world benchmarking dataset, containing numerous multimodal human-written queries. Similarly, TOUCAN \cite{TOUCAN} contains 1.5 million trajectories synthesized from nearly 500 real-world MCP servers. It can be used to evaluate agents on complex tasks, such as BFCL V3 and MCP-Universe \cite{TOUCAN}. 
Nonetheless, the communication-specific tool benchmarking dataset remains unsolved.

\begin{figure*}[tbp!]
  \centering
  \includegraphics[width=0.8\textwidth]{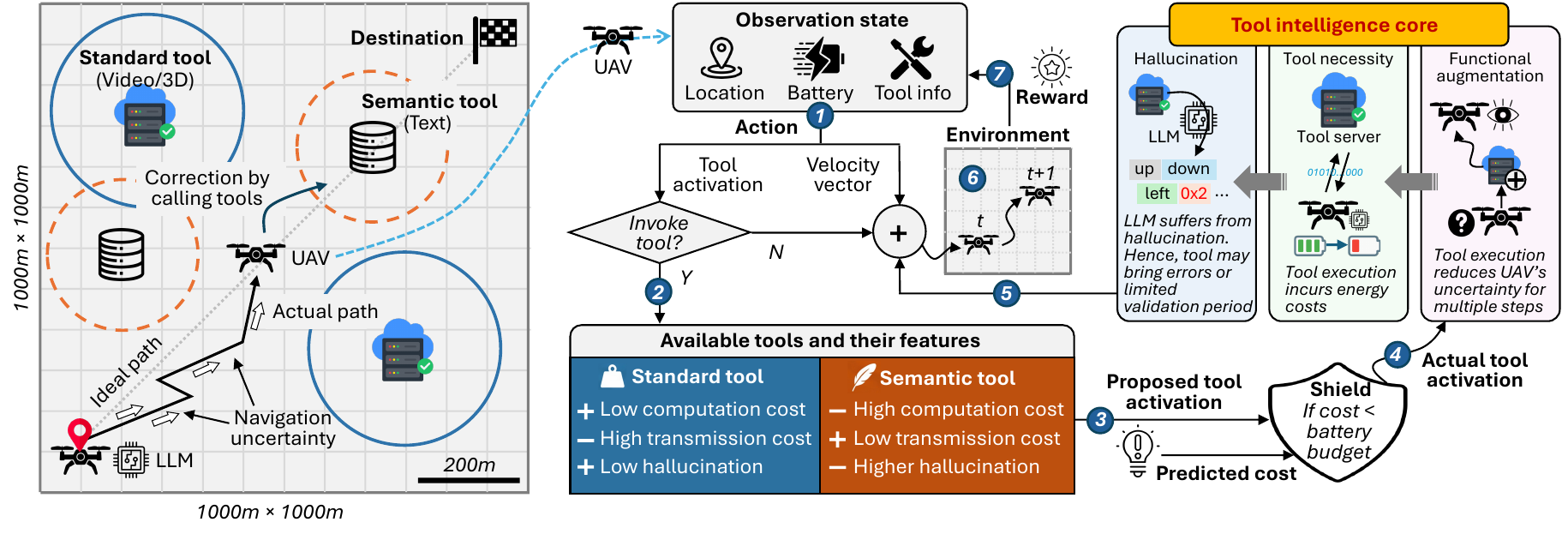} 
  \vspace{-0.3cm}
  \caption{(left): The case study scenario. (right): The procedure of the proposed algorithm. \ding{182}: The action space; \ding{183}: The illustrations of heterogeneous tools; \ding{184}: The proposed tool activation; \ding{185}: The proposed shield for training; \ding{186}: The three characteristics of tool intelligence that effects problem formulation; \ding{187}: The UAV motion; \ding{188}: The resulting reward.}
  \vspace{-0.4cm}
  \label{Fig3}
\end{figure*}
\section{Case Study}
\label{sec:case_study}

\subsection{System Model}
\subsubsection{System Overview}
UAVs in next-generation communications are evolving from simple data relays to intelligent agents capable of orchestrating various edge services. 
As shown in Fig. \ref{Fig3}, we consider that a UAV acting as a flying agent, equipped with an on-board LLM, is tasked with navigating from a starting point to a destination. 
Due to environmental complexities (e.g., fog and signal jamming) and limited sensing range, the UAV lacks accurate global situational awareness and suffers from navigational uncertainty. 

\subsubsection{Tool Settings}
To overcome these limitations, the UAV should leverage external intelligence. We consider two distinct categories of tools deployed on ground servers.
\begin{itemize}
    \item \textbf{Standard Tool:} The UAV streams raw, high-volume sensor data (e.g., video feeds) to the server. The server processes this input to reconstruct a high-fidelity 3D map for navigation. Such functionalities can be built using \textit{DroneDeploy}\footnote{https://www.dronedeploy.com/} or \textit{Google Earth Engine}\footnote{https://earthengine.google.com/} APIs. 
    \item \textbf{Semantic Tool:} The UAV sends lightweight text queries (e.g., coordinates) to the server. The server calls \textit{OpenStreetMap Overpass}\footnote{https://github.com/hrbrmstr/overpass} APIs and transmits compact semantic metadata (e.g., environmental attributes such as building height and obstacles). Crucially, the UAV employs its onboard LLM to reason over these textual descriptors to reconstruct the navigation path locally.
\end{itemize}
These APIs are encapsulated by MCP \cite{mcpzero} to form standardized agentic tools.
We suppose that the UAV employs the active tool discovery mechanism described in Section III-C, which can activate available tools within a certain range.  
To facilitate further research, we open-source a tutorial on wrapping these third-party APIs into MCP-compliant tools\footnote{https://github.com/Lancelot1998/Agentic\_AI\_Tool\_Magazine}.

\vspace{-0.3cm}
\subsection{Problem Formulation}
We consider a tool-assisted trajectory planning problem. The decision variables at each step include the UAV's flight velocity vector and a binary tool activation signal. 
Specifically, the incorporation of tool intelligence reshapes the optimization formulation, necessitating the consideration of the following critical characteristics (see Fig. \ref{Fig3}(right)).
\subsubsection{Functional Augmentation} 
Tool execution helps the UAV eliminate navigation uncertainty. This functional augmentation exerts a long-term temporal influence: a single tool activation can effectively reshape the trajectory evolution over multiple subsequent time steps, rather than merely delivering an instantaneous reward in conventional offloading.

\subsubsection{Tool Necessity} 
Tool execution incurs resource overhead. The UAV consumes significant energy to transmit data (for standard tools) or perform LLM inferences (for semantic tools), creating a coupling between communication bandwidth and computational resources. Therefore, the agent must learn to refrain from tool invocation when the tools are far away or the remaining energy is critically constrained, thus ensuring mission survival under strict resource budgets.

\subsubsection{Hallucination} 
The primary role of tools is to support the LLM's reasoning with external information. However, since LLMs function via probabilistic token generation rather than deterministic logic, they introduce an intrinsic risk of hallucination. This risk is inversely correlated with the information richness provided by the tool. Standard tools that transmit uncompressed data provide richer information, resulting in low hallucination rates. In contrast, semantic tools provide limited or abstract information. Hence, LLMs must perform heavy reasoning, significantly increasing the probability of generating factually incorrect guidance.

The functional augmentation and the associated resource overhead of tools construct a fundamental trade-off.
Moreover, hallucination imposes a limited validity horizon on the acquired guidance. 
The agent may need to perform repeated tool activations to maintain navigational accuracy throughout the flight.
Therefore, we maximize mission efficiency, which entails jointly minimizing total flight time and energy consumption (including flight propulsion, wireless data transmission, and LLM computation onboard). 
Accordingly, the optimization is subject to a constraint: the UAV's energy must never be depleted before mission completion. 
\begin{figure}[tbp!]
  \centering
  \includegraphics[width=0.3\textwidth]{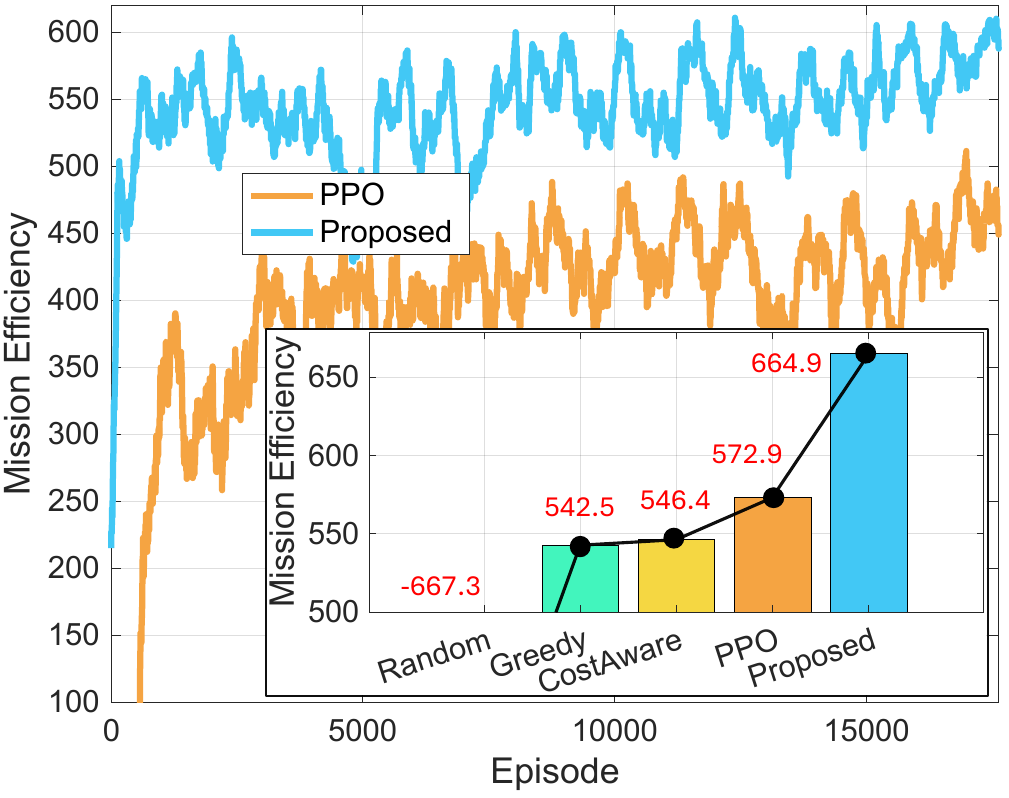} 
  \vspace{-0.3cm}
  \caption{The training curves of vanilla PPO and the proposed algorithm, and the mission efficiency of five methods.}
  \vspace{-0.6cm}
  \label{Exp_1}
\end{figure}
\begin{figure*}[htbp!]
  \centering
  \includegraphics[width=0.7\textwidth]{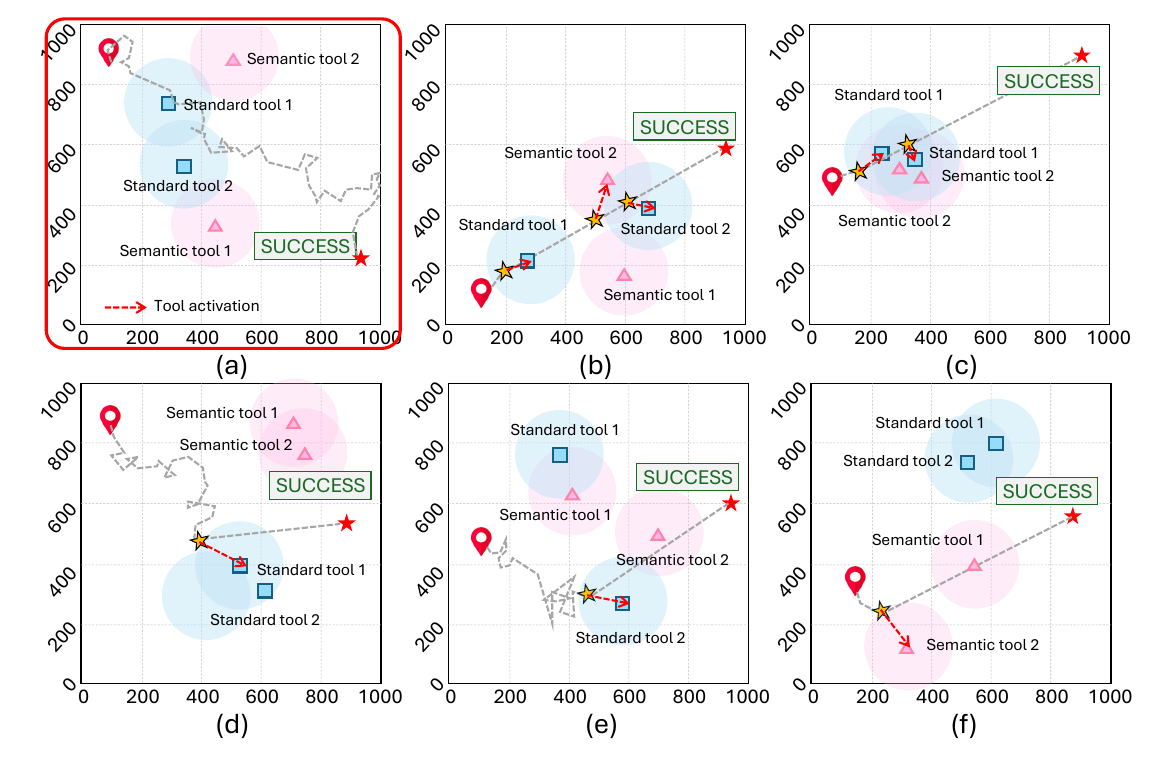} 
  \vspace{-0.3cm}
  \caption{The illustrations of UAV trajectories and tool activations. (a): The trajectory of the UAV without tools. (b)-(f): The trajectories and tool activations of the UAV trained by the proposed algorithm.}
  \vspace{-0.5cm}
  \label{Exp_2}
\end{figure*}

\vspace{-0.6cm}
\subsection{Teacher-Guided Reinforcement Learning}

We use proximal policy optimization (PPO) \cite{10582827} to train the agent. 
Moreover, to help agents efficiently operate tools while flying, we introduce a teacher shield during training.
The shield explicitly predicts the energy cost of a proposed tool call by calculating the underlying communication and computation expenditure (see Fig. \ref{Fig3}).
If the predicted cost exceeds the budget, the shield overrides the agent's action with a safe alternative, i.e., continuing to fly.
Moreover, it shapes the training process by applying a small penalty to the reward. 
This mechanism transforms sparse delayed failure signals, such as crashing after many steps, into immediate and dense feedback. 
The agent learns to associate low-battery states with the negative consequences of activating tools and internalizes the underlying physical constraints. 
Consequently, the trained agent only activates tools when necessary\footnote{This DRL agent can also be deployed as an external tool deployed in the edge network. The procedure of packing this algorithm by MCP can also be found in https://github.com/Lancelot1998/Agentic\_AI\_Tool\_Magazine.}.

\vspace{-0.3cm}
\subsection{Experimental Results}

\subsubsection{Experimental Settings}
We simulate a service area and randomly deploy four MCP servers (two with standard tools and two with semantic tools). 
To simulate the aforementioned navigational uncertainty, we inject continuous Gaussian noise into the UAV's motion dynamics at each time step. 
Without external correction, this noise accumulates, leading to significant trajectory drift.
The activations of standard/semantic tools serve as discrete state corrections of the UAV's MDP process that eliminate the deviation. 
Crucially, to capture the impact of hallucination, we set different performances for different tools. 
Specifically, standard tools can provide long-term stable navigation. 
In contrast, reflecting the higher hallucination rate, semantic tools are restricted to a shorter validity horizon.

We compare our proposed teacher-guided PPO against four distinct baselines: 1) Vanilla PPO; 2) Random Policy; 3) Greedy Policy, where the UAV navigates toward the destination and attempts to activate any accessible tool; and 4) Cost-Aware Heuristic, where the UAV navigates directly but only activates tools if the remaining energy is enough.

\subsubsection{Performance Analysis}
Here, we analyze the performance of the proposed algorithms in tool-assisted trajectory planning from three perspectives.

\textbf{Mission Efficiency:} 
As illustrated in Fig. \ref{Exp_1}, the training curves demonstrate that the proposed method significantly outperforms all baselines. 
The random policy fails to complete missions effectively, producing the worst performance. The greedy policy suffers from frequent energy depletion caused by reckless tool activations, resulting in a suboptimal reward of 542.5. The cost-aware heuristic performs competitively by avoiding fatal crashes. Finally, the proposed teacher-guided PPO achieves rapid convergence and the highest stable reward of 664.9, validating its ability to dynamically balance functional augmentation and energy costs of tools.

\textbf{Tool Activation Accuracy:} 
Figs. \ref{Exp_2}(b)-(f) reveal that the agent evolves context-aware trajectory planning and tool activation behaviors compared with the case without tools (Fig. \ref{Exp_2}(a)). 
As observed in Figs. \ref{Exp_2}(d) and (e), the agent proactively deviates from its course to enter MCP coverage, specifically when navigational uncertainty accumulates. 
Crucially, the agent further optimizes its spatial entry strategy according to the specific physics of the tool type. 
When targeting a standard tool (e.g., standard tool 1 in Fig. \ref{Exp_2}(b)), the UAV flies deep into the cell center, learning that calling standard tools costs distance-dependent transmission power.
Conversely, for a semantic tool (e.g., semantic tool 2 in Fig. \ref{Exp_2}(b)), the UAV adopts an edge skimming strategy. 
Since semantic transmission costs are distance-independent, this strategy minimizes flight energy consumption while still securing the necessary guidance.

\textbf{Tool Necessity:} 
Finally, we evaluate whether the agent achieves the cognitive ability to refrain from tool activations when they are redundant or unsafe. 
Regarding redundant invocations, Figs. \ref{Exp_2}(c) and (f) illustrate that the agent correctly minimizes unnecessary tool usage. 
Even when encountering overlapping tool coverage or passing through a tool's range while already on a low-uncertainty trajectory, the agent refrains from activation. 
This demonstrates a learned judgment that the marginal utility of further guidance is negligible compared to the activation cost. 

\vspace{-0.2cm}
\section{Future Directions}

\textbf{Zero-Trust Tool Execution Environments}:
As agents dynamically discover and invoke third-party tools in open 6G networks, ensuring operational safety becomes paramount. Future research must pivot toward establishing zero-trust tool execution environments. This involves developing cryptographic mechanisms to verify tool provenance and integrity to prevent tool poisoning.


\textbf{Semantic and Latent Tool Interfaces}:
To meet the stringent latency requirements of beyond 6G communications, tool engineering must evolve beyond current text-based interfaces, which incur significant computational overhead. Semantic and latent tool interfaces can be explored, where agents bypass high-dimensional natural language to interact directly through compressed semantics or abstracted latent representations.

\vspace{-0.3cm}
\section{Conclusion}
In this article, we have explored the transformative potential of tool intelligence in empowering agentic AI for next-generation communication networks. 
First, we have introduced a comprehensive landscape for tool engineering, detailing the essential lifecycle from tool creation and discovery to selection, learning, and benchmarking.
Moreover, we have conducted a case study about tool-assisted UAV trajectory planning, optimizing the agent's ability to activate tools while strictly adhering to the strict energy constraints. 

\bibliographystyle{IEEEtran}
\bibliography{Ref}

\end{document}